\documentclass[twocolumn,apl,amsmath,amssymb,showpacs,superscriptaddress]{revtex4-2}
\usepackage{epsf} 
\usepackage{graphicx}
\usepackage{color}
\usepackage{soul}
\usepackage{gensymb}
\usepackage{sidecap}
\usepackage{amsmath}
\usepackage{mathtools}
\DeclareUnicodeCharacter{2212}{\textendash}

\begin{document}
\title{Diffusion mechanism and electrochemical investigation of 1T phase Al-MoS$_{2}$@rGO nano-composite as a high-performance anode for sodium-ion batteries}
\author{Manish Kr. Singh}
\email{These authors contributed equally to this work}
\affiliation{Department of Physics, Indian Institute of Technology Delhi, Hauz Khas, New Delhi-110016, India}
\author{Jayashree Pati}
\email{These authors contributed equally to this work}
\affiliation{Department of Physics, Indian Institute of Technology Delhi, Hauz Khas, New Delhi-110016, India}
\author{Deepak Seth}
\affiliation{Renewable Energy and Chemicals Group, Department of Chemical Engineering, Indian Institute of Technology Delhi, Hauz Khas, New Delhi-110016, India}
\author{Jagdees Prasad}
\affiliation{Department of Physics, Indian Institute of Technology Delhi, Hauz Khas, New Delhi-110016, India}
\author{Manish Agarwal}
\affiliation{Computer Service Center, Indian Institute of Technology Delhi, Hauz Khas, New Delhi- 110016 India}
\author{M. Ali Haider}
\affiliation{Renewable Energy and Chemicals Group, Department of Chemical Engineering, Indian Institute of Technology Delhi, Hauz Khas, New Delhi-110016, India}
\author{Jeng-Kuei Chang}
\affiliation{\mbox{Department of Materials Science and Engineering, National Yang Ming Chiao Tung University, Hsinchu 30010 Taiwan}}
\author{Rajendra S. Dhaka}
\email{rsdhaka@physics.iitd.ac.in}
\affiliation{Department of Physics, Indian Institute of Technology Delhi, Hauz Khas, New Delhi-110016, India}

\date{\today} 

\begin{abstract}

We report the electrochemical investigation of 5\% Al doped MoS$_2$@rGO composite as a high-performance anode for sodium (Na)-ion batteries. The x-ray diffraction (XRD), Raman spectroscopy and high-resolution transmission electron microscopy characterizations reveal that the Al doping increase the interlayer spacing of (002) plane of MoS$_2$ nanosheets and form a stable 1T phase. The galvanostatic charge-discharge measurements show the specific capacity stable around 450, 400, 350, 300 and 200 mAhg$^{-1}$ at current densities of 0.05, 0.1, 0.3, 0.5 and 1~Ag$^{-1}$, respectively. Also, we observe the capacity retentions of 86\% and 66\% at 0.1 and 0.3 Ag$^{-1}$, respectively, over 200 cycles with a consistent Coulombic efficiency of nearly 100\%. The cyclic voltammetry, galvanostatic intermittent titration technique, and electrochemical impedance spectroscopy are used to find the kinetic behavior and the obtained value of diffusion coefficient falls in the range of 10$^{−10}$ to 10$^{−12}$ cm$^2$s$^{−1}$. Intriguingly, the {\it in-situ} EIS also explains the electrochemical kinetics of the electrode at different charge-discharge states with the variation of charge transfer resistance. Moreover, the post cycling investigation using {\it ex-situ} XRD and photoemission spectroscopy indicate the coexistence of 1T/2H phase and field-emission scanning electron microscopy confirm the stable morphology after 500 cycles. Also, the Na-ion transport properties are calculated for 1T Al--MoS$_2$@rGO interface and Al--MoS$_2$--MoS$_2$ interlayer host structure by theoretical calculations using density functional theory.  \\
\textbf{Key words:} sodium-ion batteries; anode materials; Al--MoS$_2$@rGO; electrochemical performance.

\end{abstract}

\maketitle

\section{\noindent ~Introduction}

The gradual depletion of fossil fuels and emission of greenhouse gases due to their combustion, pave the way toward renewable energy resources like solar and wind \cite{AG_RECC22,DB_nat_com19,DT_nat_com21}. However, the conversion of energy through these resources depend on several environmental factors and not ample to meet the energy demands when required. Therefore, the present research is geared towards energy storage and conversion devices for the restoration of renewable energy. As we know, the lithium-ion batteries (LIBs) are one of the most efficient technology among various energy storage devices on the subject of energy density and power density \cite{T-JES21, Y-ESM21, J-nature2001}. The LIBs are virtually the heart of all portable devices such as laptops, mobiles, as well as the electric vehicles \cite{i1, i2, Hu_EES_21, Hu_EM_20}. However, due to the increasing demand with time, and the uneven geographical and limited distribution of lithium across the world, the LIBs are becoming more expansive and not affordable for common society \cite{i3, i4}. In this context, sodium being widely and evenly distributed throughout as well as its similar electrochemistry make the sodium-ion batteries (SIBs) potential candidates for cost-effective and complementing the LIBs \cite{K-ACSEL20, HwangCSR17}. However, it is not that easy as there are great challenges, mainly coming from the larger size of sodium (1.02~\AA), which need to be resolved in order to design efficient electrode materials with appealing characteristics for energy storage applications \cite{EG_AEM20, i6, Simran21, PatiJMCA22}. Among the bottlenecks, searching the anode materials with high capacity, better rate capability, and excellent cycle life are major challenges, because graphite cannot be used due to its reactivity/thermodynamic unstability with sodium \cite{H-RCS2017,KN_JPS13}. 

At present, the SIB anode materials based on alloying and conversion are widely used because of their high theoretical capacity \cite{Liu_EEM_22, Wang_EM_22}. On the other hand, the insertion based materials provide excellent cycle life, but with a low specific capacity in SIBs \cite{MudgalPINSA22}. The sodium-ion uptake is limited for insertion based compounds due to their rigid frameworks \cite{ChandraEA20}. However, the metals and metalloids based materials have the ability of multiple sodium-ion uptakes per single atom resulting in capacities ranging from 300 to 2000 mAhg$^{-1}$ with operational voltages below 1~V vs Na/Na$^{+}$ as anode in SIBs. Mostly, group-15 (pnictogen), group-16 (chalcogen), and transition metals are used for the production of alloying and conversion based complexes \cite{HwangCSR17}. Among these, the transition metal dichalcogenides (TMDs) and particularly sulfides are of much importance because of their low cost, high capacity, and environment friendly nature. Furthermore, the low activation energy between transition metal and sulfur in TMDs facilitates the Na-ion migration during charging-discharging, while its high theoretical capacity offers high energy density for SIBs \cite{E-JPC15}. Particularly, the molybdenum disulfide (MoS$_2$) with a two-dimensional open framework has attracted considerable attention among TMDs due to its structural flexibility and high theoretical capacity as an anode material for both LIBs and SIBs \cite{ii6} as well as Zn-ion batteries \cite{Li_ACIE_21, Li_AM_21}. The MoS$_2$ is found in three phases namely 1T, 2H, and 3R where the 1T phase shows metallic nature with better ion and electron transport as compared to 3R and 2H phases because of two reasons \cite{i7,1,2,3,4,5,6,7}: (i) it has a distorted octahedral coordination structure which results in high electronic conductivity as compared to 2H and 3R phases, (ii) another advantage is its high hydrophilic nature which enables affinity of electrolytes \cite{X-nat.com16,M-Nat.nano15}. In addition to these factors, its revealing electrochemical active sites and large interlayer spacing (0.93~nm) with wide and fast ion-diffusion channels make the 1T phase more suitable for Na$^+$ ion storage \cite{4}. 

In this line, one of the most effective ways to enhance the electrochemical activity of MoS$_2$ as an anode in SIBs, is to fabricate nano-composite of MoS$_2$ particles through different synthesis routes to minimize the diffusion length of Na-ion during sodiation/de-sodiation \cite{T-SR15, C-NE17}. Chen {\it et al.} developed a scalable chemical vapor deposition method to prepare MoS$_2$ deposited electrospun carbon nanofiber hybrid to enhance the ionic conductivity, that provides large contact area for electrolyte and prevent the MoS$_{2}$ nanosheets agglomeration. This anode material exhibited a reversible capacity of 198~mAhg$^{-1}$ after 500 cycles at a current density of 1~Ag$^{-1}$ \cite{ChenActa16}. In order to study the effect of conductive carbon matrices, Sahu {\it et al.} synthesized modified 3-D framework of MoS$_2$@rGO hybrid through hydrothermal route, which was observed to deliver high discharge capacities of 588~mAhg$^{-1}$ and 501~mAhg$^{-1}$ at current densities of 100 and 500 mAg$^{-1}$ with capacity retention of 98\% and 92.3\% after 80 and 250 cycles, respectively \cite{SahuJMCA17}. Here, the rGO nano-sheets act as backbone to hold the MoS$_2$ nano-plates during cracking of the material owing to its high surface area and mechanical strength \cite{SahuJMCA17}. Also, one way to enhance the electronic conductivity and to adhere the volume expansion, is to modify MoS$_2$ with different conductive carbon matrices like rGO, CNT, carbon-doped with nitrogen and sulfur, etc. Additionally, the hetero-atom doping is also another efficient approach for improved structural stability and boosting volume expansion in MoS$_2$ \cite{7,8,9}. Li {\it et al.} proposed the improved cycling stability of Co-doped 1T MoS$_{2}$ than the pristine MoS$_{2}$, where the introduction of Co mitigates the volume expansion up to 52\% during cycling \cite{5}. Interestingly, a moderate Sn-doped 1T--2H MoS$_{2}$ anode was reported by Zeng {\it et al.}, which provides a significant rate capability of 167 mAhg$^{-1}$ at a current rate of 15 Ag$^{-1}$ with cycling retention of 320 mAhg$^{-1}$ after 500 cycles at 1 Ag$^{-1}$ \cite{ZengCC19}. Moreover, using first-principles calculations the sodium-ion intercalation and diffusion mechanism are reported in different phases of MoS$_2$/graphene layer \cite{S4}. 

Here, we have synthesized MoS$_2$, MoS$_2$@rGO, CNT-MoS$_2$@rGO, and 5\% Al doped MoS$_2$@rGO composites by a simple hydrothermal method. We found that the doping of Al ions enhance the interlayer spacing of (002) crystal plane of pure MoS$_2$@rGO nano-sheets and stabilize in a stable 1T phase, as confirmed by x-ray diffraction (XRD), Raman spectroscopy and transmission electron microscopy (TEM) measurements. Interestingly, we observe significant enhancement in the electrochemical performance of Al doped MoS$_2$@rGO, which exhibits the discharge specific capacity of around 400~mAhg$^{-1}$ at a current density of 100~mAg$^{-1}$, which found to be highly stable for more than 200 cycles even at faster charge-discharge rates with about 100\% Coulombic efficiency. The diffusive behavior in the electrode material is also investigated through detailed analysis of CV, EIS, and GITT data, and the diffusion coefficient was found to be in the range of 10$^{−10}$ to 10$^{−12}$ cm$^2$s$^{−1}$. The post-cycling analysis performed after 500 cycles at a current density of 500~mAg$^{-1}$ through room temperature XRD and FE-SEM measurements confirm the structural evolution and morphological stability of the anode material. The photoemission results indicate the 1T phase of pristine sample and coexistence of 1T/2H hetero-structure phase for the cycled anode material. We have also employed density functional theory to understand the Na-ion migration in the inter-layered structures of 1T phase Al--MoS$_2$@rGO.   

\section{\noindent ~Methods}

\textbf{2.1.1 Synthesis of graphite oxide (GO):} The GO nano-sheets were synthesized by chemical oxidation of bulk graphite powder using a slightly modified Stauden-maier process \cite{11}. In this typical procedure, 180 ml of concentrated H$_2$SO$_4$ was gradually combined with 90 ml of HNO$_3$ in a 500 ml beaker while stirring with an ice bath. We added 10 gms of graphite powder slowly into the prepared solution with continuous stirring for 30 minutes. Then, 110 gms of KClO$_3$ was added for an extended time of 2--2.5 hrs. The ice bath was withdrawn, and the solution was left to stir at ambient temperature for the next 5 days. The obtained GO solution was first washed 7-8 times with DI water and then with 10\% solution of hydrochloric acid to remove (SO$_4$)$^{2-}$ ions and other impurities. The resulting material was centrifuged 6-7 times in purified DI water. The obtained GO powder was finally vacuum dried overnight at 80$^{\degree}$C. 

\textbf{2.1.2 Synthesis of MoS$_2$, MoS$_2$@rGO, CNT-MoS$_2$@rGO:} The pure MoS$_2$, composite of MoS$_2$@rGO and CNT-MoS$_2$@rGO were prepared by a simple one step hydrothermal method. For the preparation of pure MoS$_2$ nano-sheets, 30~mmol of thiourea [(NH$_2$)$_2$CS] and 1~mmol of ammonium molybdate tetrahydrate [(NH$_4$)$_6$Mo$_7$O$_{24}$·4H$_2$O] were homogeneously dispersed for 1~hr under the constant magnetic stirring in 40~ml of DI water. The obtained black colored homogeneous solution was transferred into a 100 mL Teflon-lined stainless steel autoclave and heated at a constant temperature of 220\degree C for 24 hrs. It was followed by cooling the system to room temperature naturally so that the obtained black powder can be washed with absolute ethanol and DI water several times. Then, the obtained precipitate was dried in a vacuum oven at 65\degree C for 12 hrs and the resultant powder of MoS$_2$ was obtained. For the synthesis of MoS$_2$@rGO and  CNT-MoS$_2$@rGO composites, we use 140 mg of GO powder and 100 mg of CNT, which were dispersed homogeneously in 40 ml of DI water with the help of sonication for 1.5 hrs. After that, all the other steps were similar to the formation of pure MoS$_2$. 

\textbf{2.1.3 Synthesis of Al-doped MoS$_2$@rGO:} The Al--MoS$_2$@rGO nano-hybrids were synthesized using a facile hydrothermal method. First we mix 140 mg (3.5 mg/mL) of GO powder and 12.70 mg of AlCl$_3$.6H$_2$O in 40 mL DI water under steady ultra-sonication for 2 hrs at room temperature to make a homogenous suspended solution. After that, 30 mmol (NH$_2$)$_2$CS and 1 mmol (NH$_4$)$_6$Mo$_7$O$_{24}$·4H$_2$O were slowly added to the above formed suspension along with constant stirring for 1 hr. A homogenous dark black solution was obtained, which then placed into a Teflon-lined stainless-steel autoclave (100 mL). A box furnace was used to heat the autoclave at 220$^{\degree}$C for 24 hrs, and then cooled to room temperature when the reaction time was completed. To remove the residual ions, the obtained black precipitate was centrifuged at 3000 rpm for at least 4-5 times using DI water and absolute ethanol. Finally, a black powder of Al--MoS$_2$@rGO was obtained by drying the washed product in a vacuum oven at 65$^{\degree}$C for 12 hrs.

\textbf{2.2 Structural and physical characterizations:} To investigate the crystallographic structure of the active material, we perform the x-ray diffraction (XRD) measurements (Panalytical Xpert 3) with CuK$\alpha$ radiation (1.5406\AA) in 2$\theta$ range of 5$\degree$ to 80$\degree$. We use high-resolution transmission electron microscopy (HR-TEM) with a Tecnai G2-20 system, and field emission scanning electron microscope (FE-SEM) and energy dispersive x-ray (EDX) using the Quanta 200 FEG system to study the morphology, structure, elemental composition, and the distribution of elements across the sample. The Raman spectra were recorded with 532~nm wavelength laser using a Renishaw inVia confocal  microscope having 2400 lines/mm grating and 10mW power on the sample. The x-ray photoelectron spectroscopy (XPS) with Al K$_{\alpha}$ source (1486.6~eV), from Thermo Fisher Scientific, was used to investigate the core-level binding energy (BE) of different elements present in the sample before and after cycling. 

\textbf{2.3 Coin-cell fabrication:} The uniform slurry was prepared by taking 5\%Al--MoS$_2$@rGO as an active material, CNT as conductive carbon, and polyvinylchloride (PVDF) binder in N-methylpyrrolidone (NMP) solvent in a weight ratio of 8:1:1. The slurry was then coated on a copper foil using a doctor blade method with active material mass loading about 1 mg cm$^{-2}$ followed by vacuum drying at 120$^{\degree}$C overnight to evaporate the excess solvent and moisture. The electrodes were cut and dried in vacuum before inserting them into the glove box (UniLab Pro SP from MBraun). The CR2032 coin cells were fabricated in Ar filled glove box under a controlled level of O$_2$ and H$_2$O (less than 0.1~ppm). The Na foil was used as the counter and reference electrode.The electrolyte used was 1 M NaPF$_6$ in a mixture of ethylene carbonate (EC) and diethyl carbonate (DEC) 1:1 (vol \%) with 5 wt \% fluoroethylene carbonate (FEC) . 

\textbf{2.4 Electrochemical measurements:} The galvanostatic charge-discharge (GCD) profiles were obtained by using a Neware battery analyzer BTS400 in voltage window 0.005--2.5~V (vs Na$^+$/Na) at different current densities. The cyclic voltammetry (CV) was conducted at different scan rates using a Biologic VMP-3 model in the same voltage range as in GCD measurements. The electrochemical impedance spectroscopy (EIS) measurements were performed using a Biologic VMP-3 model in the frequency range of 10 mHz to 100 kHz, and the {\it ac} voltage amplitude was set to 10~mV at the open circuit voltage (OCV) state of the cells.

\textbf{2.5 Theoretical calculations:}

A model MoS$_2$@rGO interface is constructed with a monolayer of MoS$_2$ in a supercell size of 5$\times$4$\times$1 and 20 formula units, which are stacked with a 5$\times$5$\times$1 supercell of rGO layer. On geometry optimization of the structure, a sodium atom was introduced at one of the stable sites. Similarly, 5\% Al doped MoS$_2$--MoS$_2$ interlayer structure is constructed. The transport of sodium-ion in these host materials interlayer is simulated using plane wave basis set code of DFT as implemented in vienna {\it ab initio} simulation package (VASP-6.2) \cite{S5}. A plane wave cut-off energy of 500~eV is considered for the basis set expansion. The electron--ion core interactions are described by gradient-corrected projector augmented wave (PAW) \cite{S6} pseudo potential using Perdew-Burke-Ernzeerhof (PBE) exchange-correlation functional \cite{S7}, to solve the Kohn-Sham equations, which estimate the ground state energy of the structures. A 2$\times$2$\times$2 gamma-centred k-point mesh is used for Brillouin zone integration. The convergence criteria for energy and force are kept at 10$^{-6}$eV and 0.05 eV/\AA, respectively, for structural optimization using the conjugate gradient algorithm. To determine the minimum energy path (MEP) for sodium-ion migration and calculation of energy barriers, the Cl-NEB (climbing image nudge elastic band) method is implemented \cite{S8, S9}. The optimized structure of Na-Al-MoS$_2$ @rGO is taken as the initial state (IS) and corresponding structure after sodium-ion migration to the next available site is taken as the final state (FS) respectively, for transition state (TS) calculations. We use eight linearly interpolated images along the MEP between IS and FS. The convergence criteria considered for TS calculations are 10$^{-5}$ eV and 0.05 eV/\AA~for energy and forces respectively. The activation energy (E$_{act}$) for sodium-ion migration is calculated as the difference in energy of the TS and IS. 

\section{\noindent ~Results and discussion}

\textbf{3.1 Structural/morphological characterization:} 

\begin{figure*}
\includegraphics[width=6.6in]{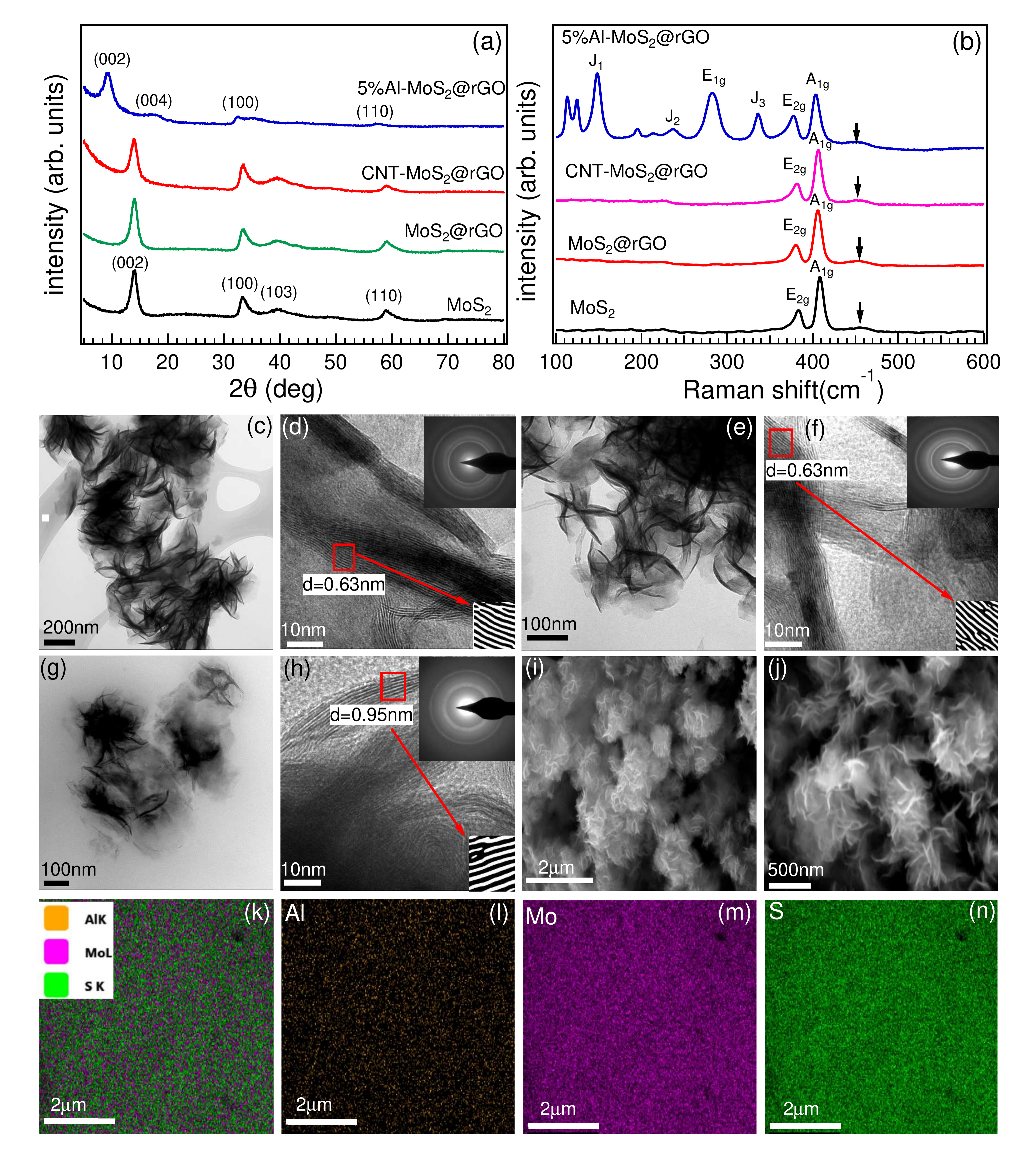}
\caption {(a) The XRD patterns and (b) the Raman spectra of MoS$_2$, MoS$_2$@rGO, CNT-MoS$_2$@rGO and Al--MoS$_2$@rGO measured at room temperature. The HR-TEM images of MoS$_2$@rGO (c, d), CNT-MoS$_2$@rGO (e, f), and Al-MoS$_2$@rGO (g, h) with the respective SAED patterns and the magnified views in the insets of (d, f, h). The FE-SEM images in (i, j), and the EDX mapping of Al, Mo and S in (k-n) for the Al-MoS$_2$@rGO composite.}
\label{ch}
\end{figure*}

First the crystal structure of as prepared MoS$_2$, MoS$_2$@rGO, CNT-MoS$_2$@rGO, and Al--MoS$_2$@rGO composites is investigated using the XRD measurements at room temperature. The diffraction patterns of MoS$_2$ MoS$_2$@rGO and CNT-MoS$_2$@rGO are appeared to be very similar to each other having four major peaks located at about 2$\theta$= 14.14\degree, 33.30\degree, 39.34\degree, and 59.06\degree, which are indexed to the lattice planes (002), (100), (103), and (110) of the crystalline 2H phase of MoS$_2$, respectively [Fig.~\ref{ch}(a)]. The peak corresponding to (002) with an interlayer spacing of about 6.3~\AA~signifies the crystalline multilayers of pure MoS$_2$ with hexagonal phase and the peak (100) depicts the stacking of Mo--S edges in vertical planes as well as the number of S active ions in the edge, as shown in Fig.~\ref{ch}(a). More importantly, the Al--MoS$_2$@rGO composite exhibits a new second order diffraction peak (004) located at about 2$\theta$ = 17.06$^{\degree}$ and the (002) peaks shifted to lower 2$\theta$ = 9.1$^{\degree}$ value where the calculated $d$ spacing found to be 5.2 and 9.3 \AA, respectively. This confirms the transformation from 2H phase to 1T phase with Al doping in MoS$_2$@rGO composite. This new phase has an increased interlayer spacing upon the intercalation of Al and NH$_3$/NH$_4$$^{+}$ ions released as by-products of thiourea used as a reductant in the hydrothermal reaction \cite{XRD_Al1, XRD_Al2, XRD_Al3}. Notably, the negatively charged surface of GO inhibits the approaching process of MoO$_{4}$$^{2-}$ precursors, which prevents the nucleation of MoS$_2$ with proper spacing between interlayers \cite{SahuJMCA17}. However, in the present case of Al doped MoS$_{2}$@rGO, the modification through Al$^{3+}$ dopant increases the electrostatic interaction between  MoO$_{4}$$^{2-}$ and negatively charged GO surfaces. This phenomenon accelerates the MoS$_{2}$ nucleation and provides large space for NH$_{4}$$^{+}$ ions occupation at the proper sites, which increases the interlayer distance along (002) direction \cite{XRD3}. For example, the presence of NH$_4$$^{+}$ ions between the layers is evident from the increment in the interlayer distance by about $\Delta$d=3 \AA, which is comparable to the size of NH$_4$$^{+}$ ions (3.75 \AA) \cite{XRD3}. Additionally, no peaks corresponding to any impurity were detected, which further indicating the phase purity of the composites. We successfully prepared 1T phase by doping 5\% Al in MoS$_2$@rGO composite having $d=$ 9.3 \AA, which is considered to be a potential anode for sodium-ion batteries \cite{XRD1, XRD2}. The Raman measurement are performed to further explore the structure of all these samples, as shown in Fig.~\ref{ch}(b). The spectra of MoS$_2$, MoS$_2$@rGO, and CNT-MoS$_2$@rGO display intense E$_{2g}$ and A$_{1g}$ peaks at about 380 and 405 cm$^{-1}$, respectively, corresponding to in-plane and out of plane vibration of S--Mo--S and peak at around 455 cm$^{-1}$ is correspond to longitudinal acoustic phonon mode \cite{raman1}. In case of the Al--MoS$_2$@rGO, the E$_{2g}$, A$_{1g}$ and peak corresponds to longitudinal acoustic phonon mode are obtained at around 377, 403 and 453.5 cm$^{-1}$, respectively. Interestingly, the specific phonon modes J$_1$ (147.8~cm$^{-1}$), J$_2$ (237~cm$^{-1}$) and J$_3$ (335.7~cm$^{-1}$) of Al--MoS$_2$@rGO corresponds to the formation of superlattice structure and symolized as the characteristic peaks of metallic 1T phase. Further, the broadening and weak intensity ratio of A$_{1g}$ to E$_{2g}$ modes also suggest for the 1T phase due to the intercalated cations as well as that the Mo is coordinated octahedrally in 1T phase \cite{Attanayake_ACSEL17}. The XRD and Raman studies demonstrate that the Al doping drives the stabilization of 1T phase in MoS$_2$@rGO composite. Also, the defects due to disorder graphitic structures are determined for these composites taking the intensity ratio of D and G bands observed in the Raman spectra, as shown in Fig.~1 of \cite{SI}. 

\begin{figure*}
\includegraphics[width=7.0in]{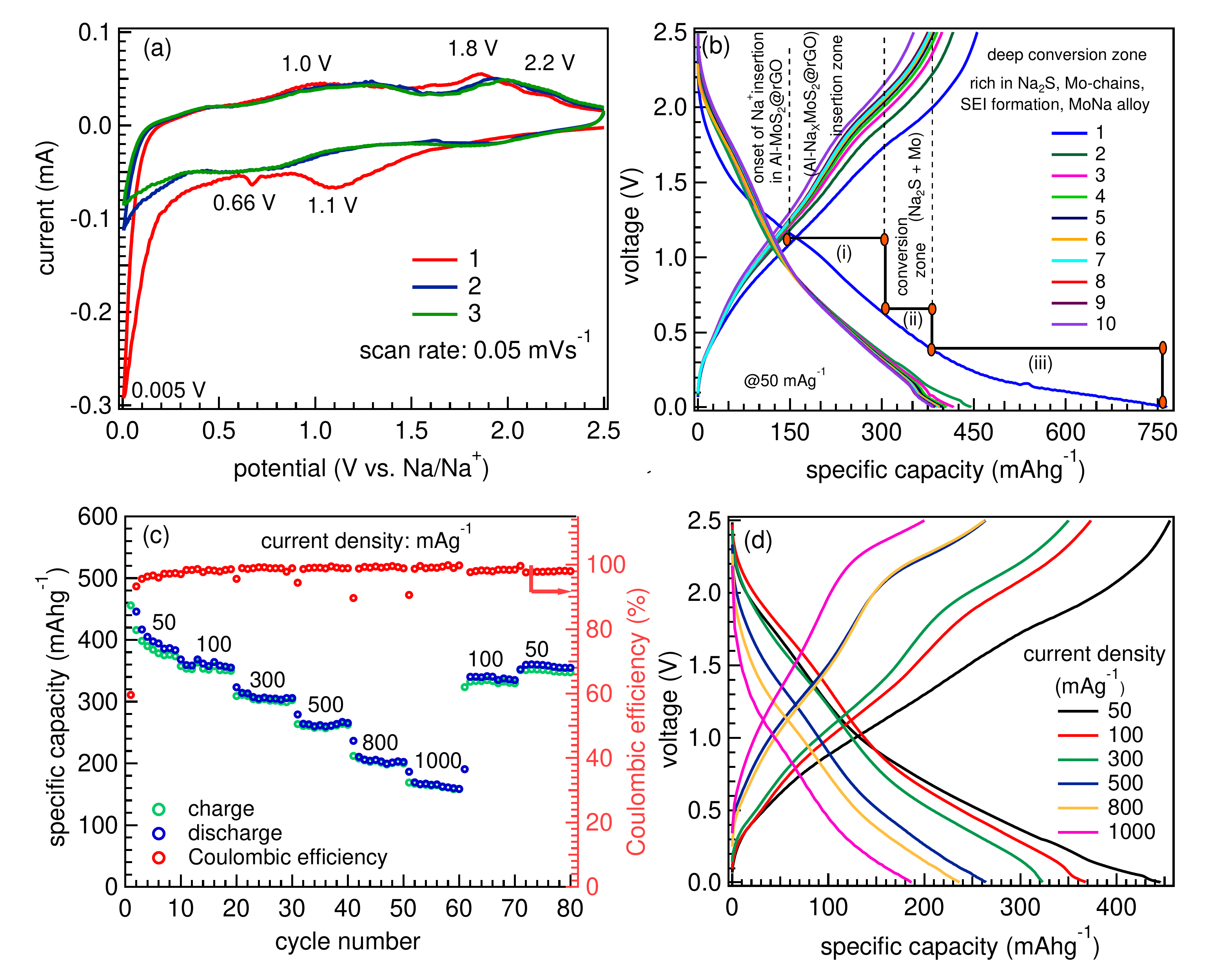}
\caption {The electrochemical measurements of Al--MoS$_2$@rGO electrode: (a) the cyclic voltammetry curves at a scan rate of 0.05~mVs$^{-1}$ for first 3 cycles in a voltage window of 0.005--2.5~V, (b) the GCD profiles at 50~mAg$^{-1}$ for the first 10 cycles, (c) the rate capability test with the Coulombic efficiency, and (d) the second cycle GCD profiles at each current densitiy.}
\label{fig2}
\end{figure*}

Figs.~\ref{ch}(c-j) show the high-resolution TEM and FE-SEM images to analyze the microscopic structure and the morphology of all the samples, MoS$_2$@rGO, CNT-MoS$_2$@rGO, and Al--MoS$_2$@rGO, respectively. In Fig.~\ref{ch}(c), the flowery architecture is clearly visible due to the 3-D interconnected MoS$_2$@rGO nano-sheets, whereas the CNT wrapped MoS$_2$ flakes are shown in the Fig.~\ref{ch}(e). These nano architectures flower like morphology show porous structures, where rGO nano-sheets are intercalated into MoS$_2$ forming nano petals. The flexible and porous rGO nano-sheets increase the electronic conductivity and active surface area of the MoS$_2$ composites due to the 3-D interconnected paths. The HR-TEM images in Figs.~\ref{ch}(d, f, h) display some ripples in the nano-sheets with few folded edges of MoS$_2$, which can be seen as the dark contrast objects well dispersed over a grey background of rGO nano-sheets. The curled edges also reveal the lattice fringes having an inter-planar spacing of 6.3~\AA, 6.3~\AA~and 9.5~\AA~corresponding to (002) crystal planes of the 2H phase of MoS$_2$@rGO and CNT-MoS$_2$@rGO, and 1T phase of Al--MoS$_2$@rGO, respectively, which are found to be consistent with the XRD patterns. The SAED patterns in the insets of Figs.~\ref{ch}(d, f, h) display hazy and broad diffraction rings. The porous nano architectures flowery like structure of Al--MoS$_2$@rGO composite with its relatively large interlayer distance provides a significant charge-storage capacity by facilitating a large number of Na-ion intercalation and also enhances the stability owing to the higher mechanical strength of rGO nano-sheets \cite{FESEM}. The FE-SEM images in Figs.~\ref{ch}(i, j) are also conforming the flower like morphology of Al--MoS$_2$@rGO composite. In Figs.~\ref{ch}(k-n), the elemental mapping images of Al--MoS$_2$@rGO confirm the uniform distribution of Al, Mo and S elements. The thermogravimetric analysis (TGA) confirms around 7wt\% of the rGO content in the sample, as shown in Fig.~2 of \cite{SI}. 

\textbf{3.2 Cyclic voltammetry (CV) and galvanostatic charge-discharge (GCD) investigation:} 

In order to demonstrate the redox mechanism during sodium ion intercalation in 1T Al--MoS$_2$@rGO host electrode, the CV measurement are performed at a scan rate of 0.05~mVs$^{-1}$ in the voltage window of 0.005--2.5~V. Fig.~\ref{fig2}(a) displays first few cycles of CV profile, which depicts three reduction (cathodic) peaks/shoulders at 1.1, 0.66 and 0.005~V and three corresponding oxidation (anodic) peaks at 1, 1.8 and 2.2~V. The first cathodic peak at 1.1~V involves the formation of Na$_x$MoS$_2$ due to the Na-ion insertion into the MoS$_2$ host structure as well as the SEI formation \cite{8, CV1}, whereas the peak at 0.66~V is assigned to the conversion of Na$_x$ MoS$_2$ resulting disintegration of MoS$_2$ into Mo particles embedding in Na$_{2}$S matrix \cite{CV3}, see for example below equations:
\begin{equation}
\mathrm{MoS}_{2}+x \mathrm{Na}^{+}+x \mathrm{e}^{-} \rightarrow \mathrm{Na}_{x} \mathrm{MoS}_{2}
\label{1}
\end{equation}
\begin{equation}
\mathrm{Na}_{x} \mathrm{MoS}_{2}+(4-x) \mathrm{Na}^{+}+(4-x) \mathrm{e}^{-} \rightarrow 2 \mathrm{Na}_{2}\mathrm{S}+\mathrm{Mo}
\label{2}
\end{equation}
Additionally, the third sharp cathodic peak at 0.005~V indicates the intercalation of Na-ion between Na$_{2}$S and Mo interface \cite{6}. The oxidation peak at 1.8~V is attributed to the reverse reaction of metallic Mo nano-grains and Na$_2$S as described below \cite{SahuJMCA17}: 
\begin{equation}
\mathrm{Mo}+2\mathrm{Na}_{2} \mathrm{S}  \rightarrow 4 \mathrm{Na}^{+}+ \mathrm{MoS}_{2} + 4\mathrm{e}^{-} 
\label{3}
\end{equation} 
whereas, a small anodic peak at 2.2~V represents the oxidation of Na$_{2}$S to S or polysulfide \cite{WangAS19}. The subsequent CV curves of 1T Al--MoS$_2$@rGO anode are almost overlapped after the 1$^{st}$ cycle, implying a reversible sodiation/desodiation process and electrochemical stability. 

\begin{figure*}
\includegraphics[width=7.2in]{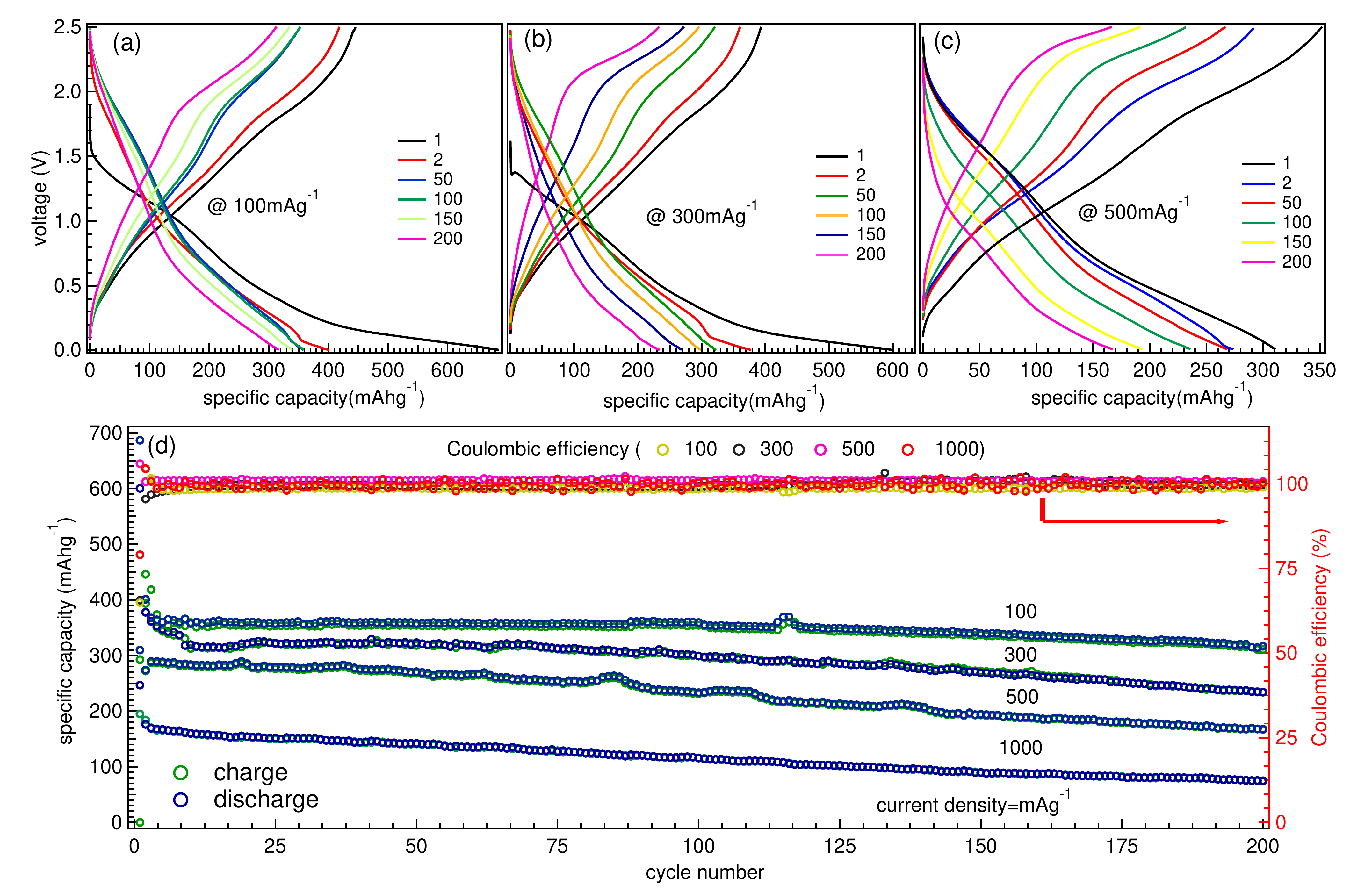}
\caption {The GCD curves of Al--MoS$_2$@rGO electrodes for selected cycles at a current density of (a) 100 mAg$^{-1}$, (b) 300 mAg$^{-1}$, and (c) 500 mAg$^{-1}$. (d) The specific capacity during long cycling up to 200 cycles at different current densities.}
\label{fig3}
\end{figure*}

Fig.~\ref{fig2}(b) shows the galvanostatic charge-discharge (GCD) curves in the voltage window of 0.005--2.5~V for first 10 cycles at current density of 50 mAg$^{-1}$. The GCD curves show sloping features in the initial cycle and gradually inconspicuous plateaus are observed in the subsequent cycles during sodiation process. Notably, three plateaus are clearly visible in the initial sodiation process, which are consistent with the CV scans, where the first plateau located around 1.1~V denotes the insertion of Na-ions into the interlayers of MoS$_2$ (as discussed in equation~1), whereas the second plateau originates at 0.6~V corresponds to the conversion reaction involving the formation of metallic Mo and Na$_2$S matrix, and the third plateau at 0.4-0.005~V represents the intercalation of Na-ion into the interface of Mo and Na$_2$S, as marked in Fig.~\ref{fig2}(b). On the other hand, the sloping region at 1.9~V is clearly visible in case of charge, representing the reversible oxidation of Mo to MoS$_2$. We find a significant enhancement in the electrochemical performance of Al--MoS$_2$@rGO composite due to its stable metallic 1T phase with a larger interlayer spacing. The first discharge specific capacity is found to be 703~mAhg$^{-1}$ at the current density of 50~mAg$^{-1}$, which is higher than the theoretical specific capacity of MoS$_2$ (670~mAhg$^{-1}$, this theoretical capacity is evaluated on the basis of conversion reaction between one MoS$_2$ molecule and four sodium ions). The stable specific capacity retention of about 400~mAhg$^{-1}$ was observed at 50~mAg$^{-1}$ in the following cycles. The additional capacity obtained in the first discharge cycle may be attributed to the formation of SEI layer at the electrode-electrolyte interface, which is caused by electrolyte reduction and also exposure to vacancy defects of rGO \cite{extra_capacity1}. 

The remarkable rate performance of 1T Al--MoS$_2$@rGO anode is presented in Figs.~\ref{fig2}(c, d). Here, we observe high rate capability at different current densities of 50, 100, 300, 500, 800 and 1000~mAg$^{-1}$, approaching 446~mAhg$^{-1}$ at 50 mAg$^{-1}$ to a specific capacity of 169~mAhg$^{-1}$ at 1000~mAg$^{-1}$ and the corresponding GCD profiles are shown in Fig.~2(d) for the second cycle at each current density. The electrodes display 91\% and 95\% specific capacity retention as the current density returns to 50 and 100 mAg$^{-1}$ indicating high stability of the formed SEI layer in FEC added electrolyte \cite{6}. In order to elucidate the long stability of the cell performance, Figs.~\ref{fig3}(a, b, c) display the GCD profiles for every 50$^{th}$ cycle at current densities of 100, 300 and 500~mAg$^{-1}$, respectively. The GCD curves in Fig.~\ref{fig3}(c) are measured on the same cell after 5 cycles at 50 mAg$^{-1}$ (shown in Fig.~3(a) of \cite{SI}). Moreover, the long cycling performance of Al--MoS$_2$@rGO anode is shown in Figs.~\ref{fig3}(d) at different current density over 200 cycles. The composite as an anode shows capacity retention of 86\%, 66\%, 58\%, and 45\% at the current densities of 100, 300, 500, and 1000~mAg$^{-1}$ with almost 100\% Coulombic efficiency, showing better stability at slower current rates. The Al doping in MoS$_2$@rGO lowers the activation barrier of Na-ion for ease insertion/extraction into the host-structure by increasing the interlayer distance, which improves the specific capacity and stability. However, the slow capacity fading during the long cycling process, as shown in Fig.~3 of \cite{SI}, may involve some polarization effect, surface stability as well as the transition-metal cation (Mo) dissolution leading the formation of a thick SEI layer \cite{HouAEL21}. Here, the particle size, surface engineering by decorating conducting materials as well as morphology control may help to further improve the cyclic stability \cite{HuAC14}.  

\textbf{3.3 Analysis of cyclic voltammetry data:}

\begin{figure*}
\includegraphics[width=7.1in]{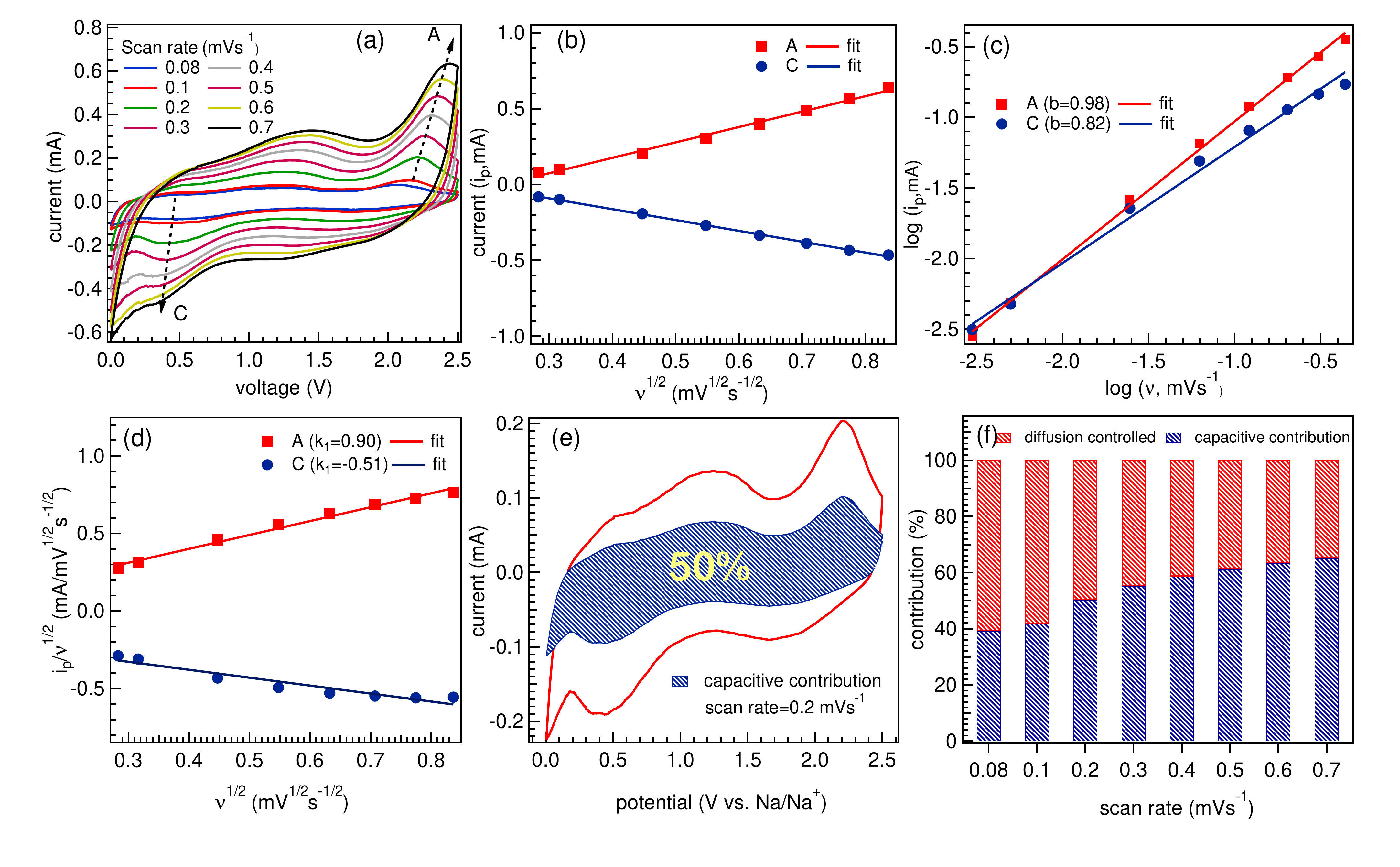}
\caption {(a) The cyclic voltammetry curves of Al--MoS$_2$@rGO at different scan rates ranging from 0.08 to 0.7 mVs$^{-1}$ in a potential window of 0.005--2.5~V, (b) the linear fit between peak current i$_p$ and square root of scan rate, and (c) the linear fit between log(i$_p$) and log($\nu$) for different scan rates, (d) the linear fit between i$_p$/$\nu$$^{1/2}$ and $\nu$$^{1/2}$ for different scan rates, (e) the shaded area showing capacitive contribution at a scan rate of 0.2~mVs$^{-1}$, and (f) the capacitive and diffusive contributions for the total current extracted at various scan rates.}
\label{3.CV-pseudo-capa}
\end{figure*}

Further, to explore the sodium-ion storage mechanism in Al--MoS$_2$@rGO anode, the kinetic behavior of the ions is studied by detailed CV measurements performed at various scan rates ranging from 0.08 to 0.7~mVs$^{-1}$ in a voltage window of 0.005--2.5~V, as presented in Fig.~\ref{3.CV-pseudo-capa}(a). The diffusion coefficient is calculated using Randles-Sevcik equation as given below \cite{D-CV}: 
\begin{equation}
i_{p}=\left(2.69 \times 10^{5}\right) A D^{\frac{1}{2}} C n^{\frac{3}{2}} \nu^{\frac{1}{2}}
\label{6}
\end{equation}
where, $i_p$ is the peak current (mA), $A$ is the active surface area of the electrode (cm$^2$), $D$ is the diffusion coefficient (cm$^2$s$^{-1}$), C is the sodium ion concentration in the bulk (mol cm$^{-3}$), $n$ is the electron transferred in the redox reaction and $\nu$ is the scan rate (mVs$^{-1}$). In order to calculate the value of $D$, we fit the oxidation/reduction peak current versus square root of $\nu$ with linear behavior, as shown in Fig.~\ref{3.CV-pseudo-capa}(b) and the slope is extracted. The $D$ values were found to be 1.60$\times$10$^{-10}$ and 0.91$\times$10$^{-10}$ cm$^2$s$^{-1}$ correspond to the anodic and cathodic peaks, respectively. In the anodic case, the relatively fast Na-ion kinetics is due to the disappearance of non-conductive Na$_{2}$S \cite{T-SR15,WangAS19}, which is also consistent with the gradual decreasing trend of resistance in the EIS spectra during the charging from 0.005 to 2.5~V [as shown later in Figs.~5(b, d)]. 

\begin{figure*}
\includegraphics[width=7.1in]{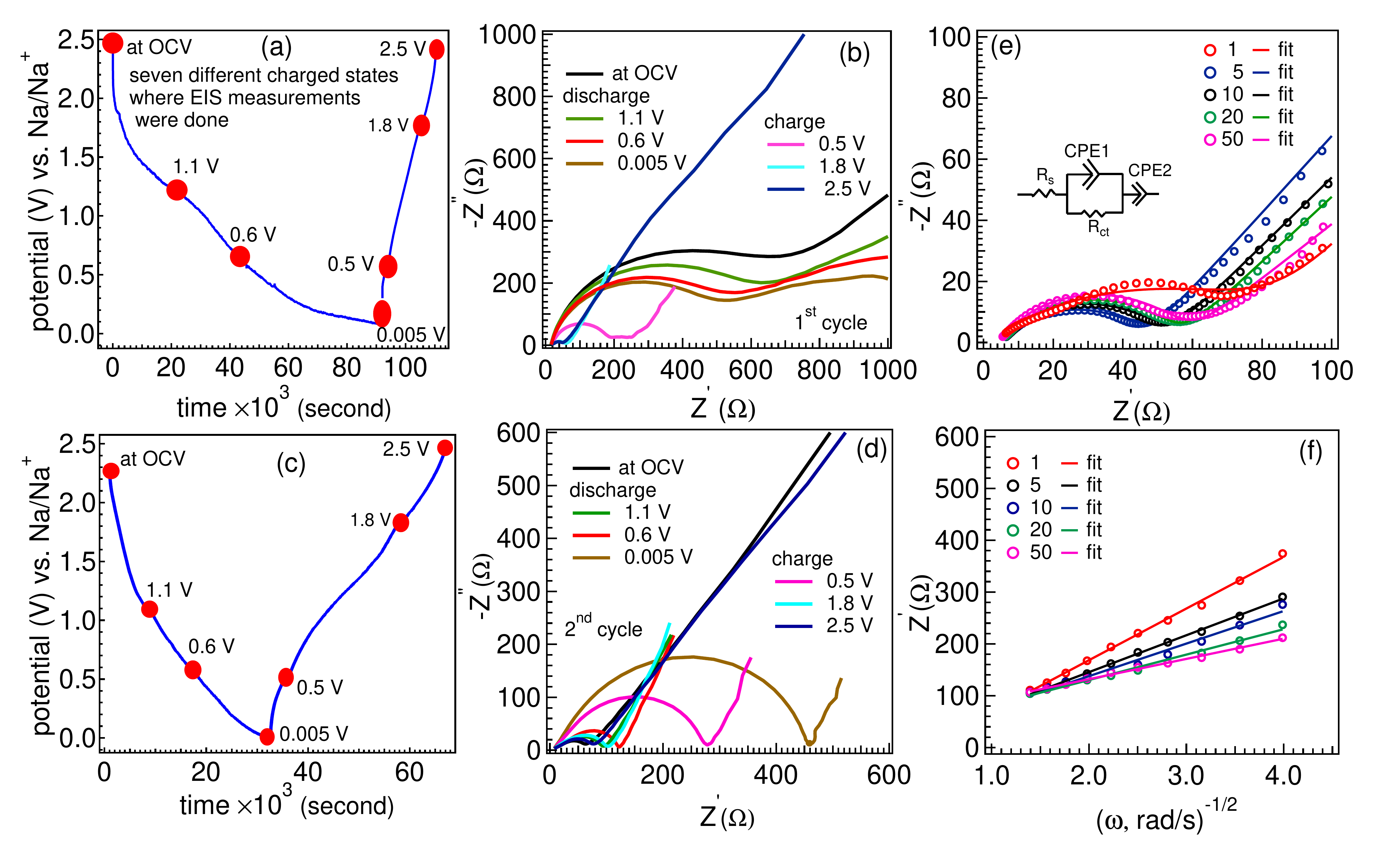}
\caption {The gavanostatic charge-discharge profile of the (a) first cycle and (c) second cycle with seven different points marked where the EIS measurements are carried out. The {\it in-situ} Nyquist plot during the (b) first cycle and (d) second cycle at different critical voltages ranging from OCV to fully sodiated/discharged state (0.005~V), and then fully desodiated/charged state (2.5~V), (e) the fitted Nyquist plots, with the corresponding equivalent circuit, measured in fully discharged states after 1$^{st}$, 5$^{th}$, 10$^{th}$, 20$^{th}$ and 50$^{th}$ cycles of GCD at 500 mAg$^{-1}$, (f) the plot of the real part of impedance versus the inverse square root of angular frequency in Warburg region at different cycles obtained from data in (e).}
\label{EIS}
\end{figure*}

Moreover, the current response of the electrode at different potentials can be attributed to the surface-controlled and diffusion-controlled currents. The net current obeys the following relation \cite{dunn1}: 
\begin{equation}
i=a \nu^{b}
\label{3}
\end{equation}
where $i$ is the net current in mA, $\nu$ (mVs$^{-1}$) is the scan rate and $a$ and $b$ are adjustable parameters. Here, we can identify different charge storage mechanisms by the value of $b$, which normally varies between 0.5 to 1; for example, $b=$ 0.5 represents a diffusion-controlled mechanism and $b=$ 1 represents a surface capacitive current. It is important to note that the value of $b$ can be obtained by the slope of liner fitting of log(peak current i$_p$) versus log($\nu$), as shown in Fig.~\ref{3.CV-pseudo-capa}(c). The obtained values of $b$ for anodic peaks (A) and cathodic peaks (C) are 0.98 and 0.82, respectively, suggesting the pseudo-capacitive behavior of the Al--MoS$_2$@rGO electrode, which shows a faradic like charge storage mechanism with capacitor like electrochemical response. This type of behavior is basically observed in nano-structured MoS$_2$, which enable the system to achieve high energy and power densities \cite{ChoiNRM20}. Moreover, based on the previously reported analysis by Dunn and his co-workers, the current response during voltage sweep can be written as the sum of both capacitive and diffusive current, as given below \cite{ii6, dunn3, dunn4}:
\begin{equation}
i=k_{1} \nu+k_{2} \nu^{1 / 2}
\label{4}
\end{equation}
After rearranging the above equation, we can write as: 
\begin{equation}
i / \nu^{1 / 2}=k_{1} \nu^{1 / 2}+k_{2}
\label{5}
\end{equation}
where, k$_1$ and k$_2$ are the adjustable parameters and k$_1$$\nu$ and k$_2$$\nu^{1/2}$ show the contribution from capacitive and diffusive currents, respectively. To calculate the value of k$_1$ and k$_2$, we perform the linear fit to the peak current $i_p/ \nu^{1 / 2}$ versus $\nu^{1 / 2}$ plot, as shown in Fig.~\ref{3.CV-pseudo-capa}(d). The slope and intercept of the line give the values of k$_1$ and k$_2$, respectively. The obtained values of k$_1$  for anodic/cathodic peaks are found to be 0.90/0.51, respectively. Using the values of k$_1$ and k$_2$, the calculated capacitive and diffusive contributions are found to be 50\% each, as plotted for the scan rate of 0.2~ mVs$^{-1}$ in Fig.~\ref{3.CV-pseudo-capa}(e). The variation in the diffusive and capacitive contributions (in percentage scale) for different scan rates is shown in Fig.~\ref{3.CV-pseudo-capa}(f), where 0.7~mVs$^{-1}$ attributes lowest diffusion controlled current. The increasing trend of capacitive contribution indicates that at higher scan rates most of the charge is being stored by the surface redox reaction. \\

\textbf{3.4 ~Electrochemical impedance spectroscopy:}

\begin{figure*}
\includegraphics[width=7.1in]{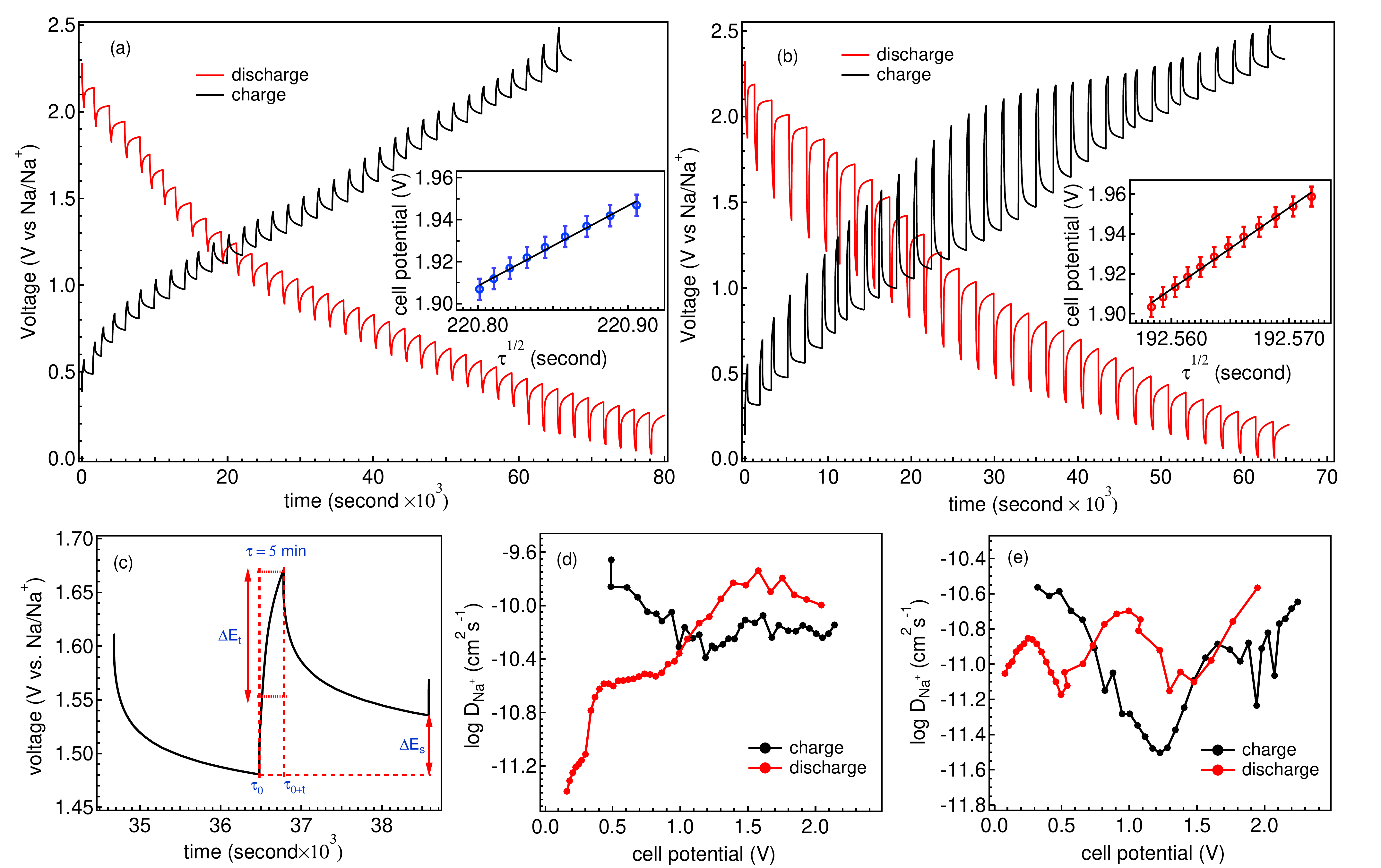}
\caption {The GITT measurements of Al--MoS$_2$@rGO electrodes at a current density of 500~mAg$^{-1}$ in a voltage window of 0.005--2.5~V (a) during 4$^{th}$ cycle, and (b) 500$^{th}$ cycle, (c) a schematic labeling of different parameters of a single titration curve before, during and after application of a current pulse for 5 min. The logarithimic diffusion-coefficient profile as a function of cell voltage during charge--discharge for 4$^{th}$ cycle in (c), and for 500$^{th}$ cycle in (d).}
\label{GITT}
\end{figure*}

To reveal the electrochemical reaction kinetics of the 1T phase Al--MoS$_2$@rGO composite, the EIS measurements are performed at room temperature in the frequency range of 100~kHz to 10~mHz at seven different states during the first sodiation/de-sodiation cycle, as marked across the potential vs time curve in Fig.~5(a). The corresponding Nyquist plot at OCV in Fig.~5(b) consists of a compressed semicircle in the high and middle range of frequency regions, which are attributed to the charge transfer resistance (R$_{\rm ct}$) and R$_{\rm SEI}$ at the electrode-electrolyte interface and the straight line in the low-frequency region is related with the diffusion of sodium-ion in the bulk region of electrode material. During the first discharge from 1.1~V to 0.005~V (as shown in Fig.~\ref{EIS}(b)), nearly two semicircles are evolved in the high and mid frequency range, depicting the SEI formation and the intercalation of sodium-ion in the MoS$_2$ host matrix. Also, the non-conductive Na$_2$S matrix due to the constant volume change of the active material occurring during discharging of the cell. The formation of nano-clusters like Na$_x$MoS$_2$ gradually decreases the charge transfer resistance during de-sodiation. However, in the case of charging, the semi-circle in the intermediate frequency range disappears and a single semi-circle formed in the entire frequency region, elucidating the dominance of charge transfer resistance over the SEI impedance \cite{ChoiJES20}. The subsequent decrement in the diameter of the semicircle, when charging from 0.5~V to 2.5~V, shows the dissolution of the thick SEI layer. Note that the first cycle is also complicated having some irreversible nature possibly due to SEI formation and electrolyte reduction, which is also reflected in Fig.~5(b) where the R$_{\rm ct}$ found to decrease with discharge/charge voltages. Therefore, to understand the reaction mechanism in more stable state, we present the similar data on a fresh cell for the 2$^{nd}$ cycle in Figs.~5(c, d), which clearly show the increasing/decreasing trend in the R$_{\rm ct}$ values during discharge/charge process confirming the reversible mechanism from second cycle. 

Furthermore, the EIS measurements are performed to study the variation in R$_{\rm ct}$ values and diffusion coefficient after several cycles where the Nyquist plots are shown in Fig.~\ref{EIS}(e). The value of the physical components corresponding to different regions can be calculated by fitting the equivalent circuit model \cite{8}, as shown in the inset of Fig.~\ref{EIS}(e) at different cycles. Here, the high value of R$_{\rm ct}$ in 1$^{st}$ cycle as compared to the 5$^{th}$ cycle is a bit unusual and not consistent with the observed specific discharge capacity during the initial cycles, as shown in Fig.~3(a) of \cite{SI}. This could be due to several reasons like complex reaction at electrode/electrolyte interface; however, a similar observation in refs.~\cite{ChenJMCA15,ChenNML19} is said to be the activation process during initial cycle(s). After the 5$^{th}$ cycle, the subsequent increase in the R$_{ct}$ value determines the in-filtration of sodium-ion and the formation of high activation barrier to hinder the sodium-ion kinetics. This phenomenon proves that the surface stability of MoS$_2$ decreases during cycling, which is consistent with the long cycling measurement of MoS$_2$, as shown in Fig.~3(d). The diffusion coefficient of sodium-ion in the bulk of Al--MoS$_2$@rGO electrode can be calculated by the following equations \cite{EIS2}:
\begin{subequations}
\begin{equation}
D_{\mathrm{Na}^{+}}=R^{2} T^{2} / 2 n^{4} F^{4} \sigma_{\mathrm{w}}^{2} A^{2} C^{2} 
\label{7} 
\end{equation}
\begin{equation}
Z^{\prime}=R_{s}+R_{c t}+\sigma \omega^{-0.5}
\label{8} 
\end{equation}
\end{subequations}
where $R$, $T$, $n$, $F$, $\sigma_w$, $A$, and $C$, are the universal gas constant (8.314 J mol$^{−1}$ $K^{-1}$), temperature, the number of electrons exchanged in the reaction, the Faraday's constant (96485 C mol$^{-1}$), the Warburg coefficient ($\ohm-s^{-1/2}$), the surface area of the electrode (1.13 cm$^2$) and the Na-ion concentration in the electrode material (mol-cm$^{-3}$), respectively. The Warburg coefficient $\sigma_w$ is related to the real part of impedance (Z$^{\prime}$) by equation~8, where R$_s$ is the solution resistance. In order to calculate the $D_{\mathrm{Na}^{+}}$, the real part of impedance is plotted as a function of the inverse square root of angular frequency $\omega^{-1/2}$ in the Warburg region at different cycles and fitted linearly as shown in Fig.~\ref{EIS}(f). The values of the apparent sodium diffusion coefficient are found to be in the range from 0.18$\times$10$^{-11}$ to 1.84$\times$10$^{-11}$ cm$^2$s$^{-1}$.

\textbf{3.5 ~Galvanostatic intermittent titration technique (GITT) investigation:}

The chronopotentiometry based GITT measurements were performed on the electrode to demonstrate the structural evolutions and the origin of the electrochemical kinetics during the Na-ion insertion/extraction process. Figs.~6(a, b) elucidate the overall potential response profiles as a function of time at 4$^{th}$ cycle and 500$^{th}$ cycle, respectively, and are consistent with the galvanostatic charge-discharge curves. The GITT measurements were carried out by first charging the cells with a current density of 500 mAg$^{-1}$ to maximum cut-off voltage for a short duration of 5 min, which was followed by a relaxation period of 30 min to reach a steady-state value (E$_{s}$), then the same condition was repeated to discharge the cells up to the minimum voltage. To calculate the diffusion coefficient under thermodynamic equilibrium conditions, the Fick's second law of diffusion is simplified by using the following equation with some assumptions \cite{JingNR19}.  
\begin{eqnarray}
D_{Na^{+}}= \frac{4}{\pi \tau}\left[\frac{m V_{\rm M}}{M A}\right]^2\left[\frac{\Delta E_s}{\tau(\Delta E_t/d\sqrt \tau)}\right]^2
\end{eqnarray}
where, $\tau$=L$^{2}$/D$_{Na^+}$, L is the diffusion length of sodium ion (thickness of the electrode), V$_{M}$ (cm$^3mol^{-1}$) is the molar volume of Al--MoS$_2$@rGO composite, which is deduced using the equation “N$_A$*V$_{unit}/Z$” (here N$_A$ is the Avogadro's number, V$_{unit}$ is the volume of a unit cell and $Z$ is the number of unit cells). The m(gm) and M(gm mol$^{-1}$) are the mass and molecular weight of Al doped MoS$_2$, respectively, $A$ is the active surface area between the electrode and electrolyte, $\Delta$E$_{s}$ and $\Delta$E$_{t}$ are the changes in steady-state voltage after the OCV stand of 30 min, and the transient change in voltage due to the current pulse after excluding the IR drop, respectively, as depicted in Fig.~6(c). Here, the molar volume is assumed to be constant with the variation of the state of charge (SOC) for diffusion coefficient calculation. The insets of Figs.~6(a, b) shows a linear relationship between the variation of cell voltage and the square root of time ($\tau$$_{1/2}$), which is used as a fact to simplify the equation~9, and that can be written as below:
\begin{eqnarray}
D_{Na^{+}}= \frac{4}{\pi \tau}\left[\frac{m_B V_M}{M_B A}\right]^2\left[\frac{\Delta E_s}{\Delta E_t}\right]^2;\tau=L^{2}/D_{Na^+}
\end{eqnarray}
The Na-ion diffusion coefficient profiles as a function of cell voltage during charge and discharge processes for 4$^{th}$ and 500$^{th}$ cycles are shown in Figs.~6(d, e), respectively. In the case of discharge process, the diffusion coefficient of Na$^{+}$ gradually decreases, showing a gradual increment of over-potential during insertion of sodium ions into Al--MoS$_2$@rGO host. However, the charging process shows a stable profile, elucidating lower potential barriers for sodium-ions. The diffusion coefficient profile for 500$^{th}$ cycle, as demonstrated in Fig.~6(e) represents slightly slow diffusion of Na-ions in the host structure, i.e., diffusion coefficient is somewhat lower than the 4$^{th}$ cycle. These results are consistent with the long cycling data in Fig.~3, where there is a gradual degradation in capacity after 200 cycles. The minimum in the profile depicts the phase transition points, leading to strong interactions between sodium-ion and the host matrix. 

\textbf{3.6 ~Post cycling XRD and FESEM:}

The post-cycling tests are carried out to investigate the effect of cycling on the structure and morphology of the extracted active material. We have dismantled the cell after 500 cycles at a current density of 500 mAg$^{-1}$ and washed the electrode with DEC to retrieve the active material. 
\begin{figure}[h]
\includegraphics[width=3.45in]{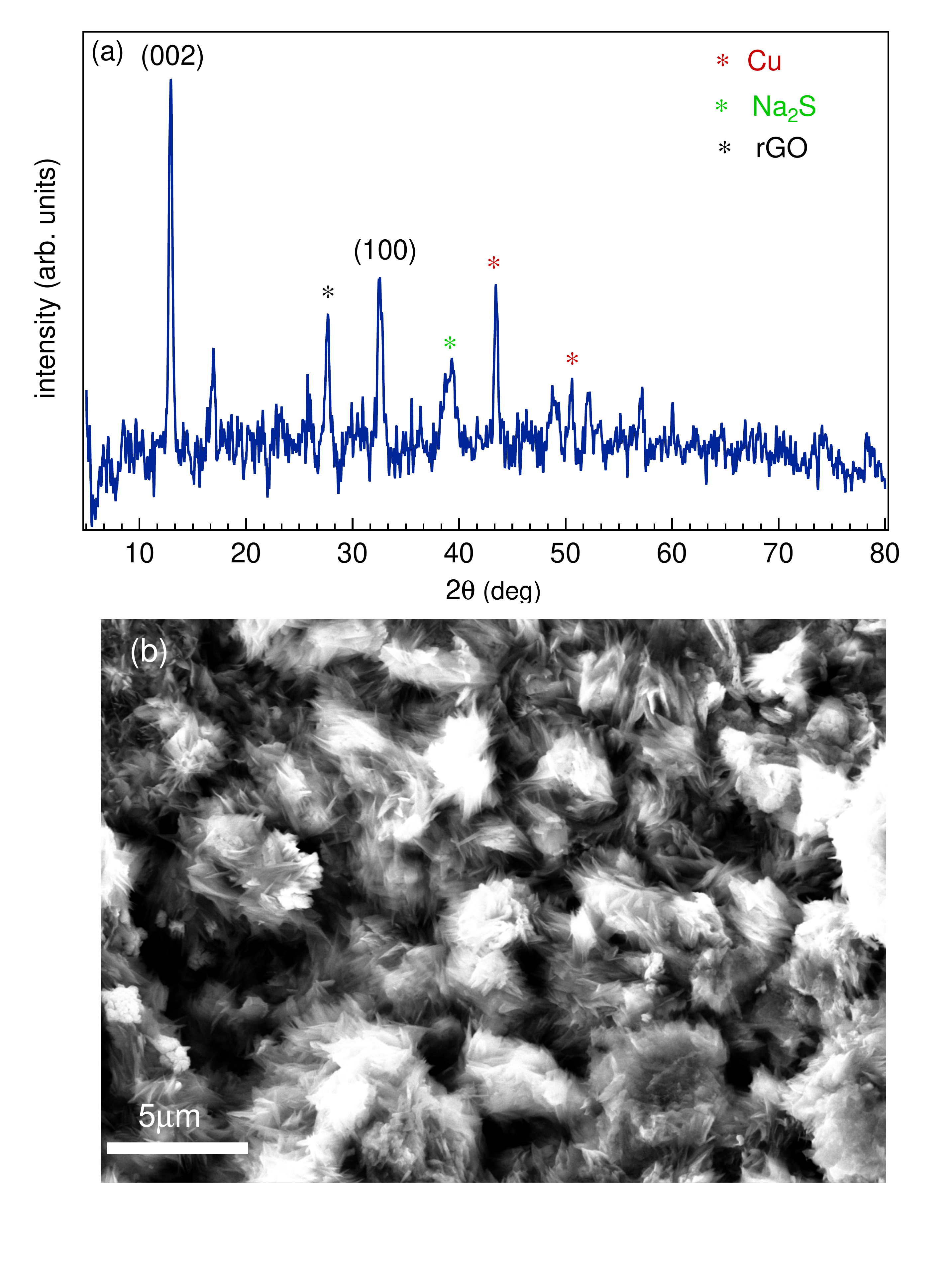}
\caption {(a) The XRD pattern and (b) the corresponding FE-SEM image of extracted active material from the cell after 500 cycles at a current density of 500~mAg$^{-1}$.}
\label{post-analysis}
\end{figure}
Fig.~\ref{post-analysis}(a) depicts the XRD pattern, which shows the peaks located at 14$^{\degree}$ and 33.3$^{\degree}$ that are attributes to the (002) and (100) planes, indicating the crystalline nature of the MoS$_2$ after cycling. The (220) peak at 39 $^\circ$ (JCPDS No.03-0933) reflects the existence of Na$_2$S due to the incomplete reaction after cycling \cite{T-SR15} and some peaks at around 43$^{\degree}$ and 51$^{\degree}$ are observed from the Cu-foil \cite{6}. The shifting of (002) peak towards higher 2$\theta$ suggests for a structural transformation of some percentage of 1T phase to 2H phase in the active material, which can be the primary cause of capacity fading after long cycling. The corresponding FE-SEM image is shown in Fig.~\ref{post-analysis}(b) where the morphology of the active material indicates no cracking and discontinuity in the structure, which attributes to the corresponding capacity retention of the composite at 500 mAg$^{-1}$ after long cycling.

\textbf{3.7 ~X-ray photoemission spectroscopy study:}

\begin{figure}[h]
\includegraphics[width=3.5in]{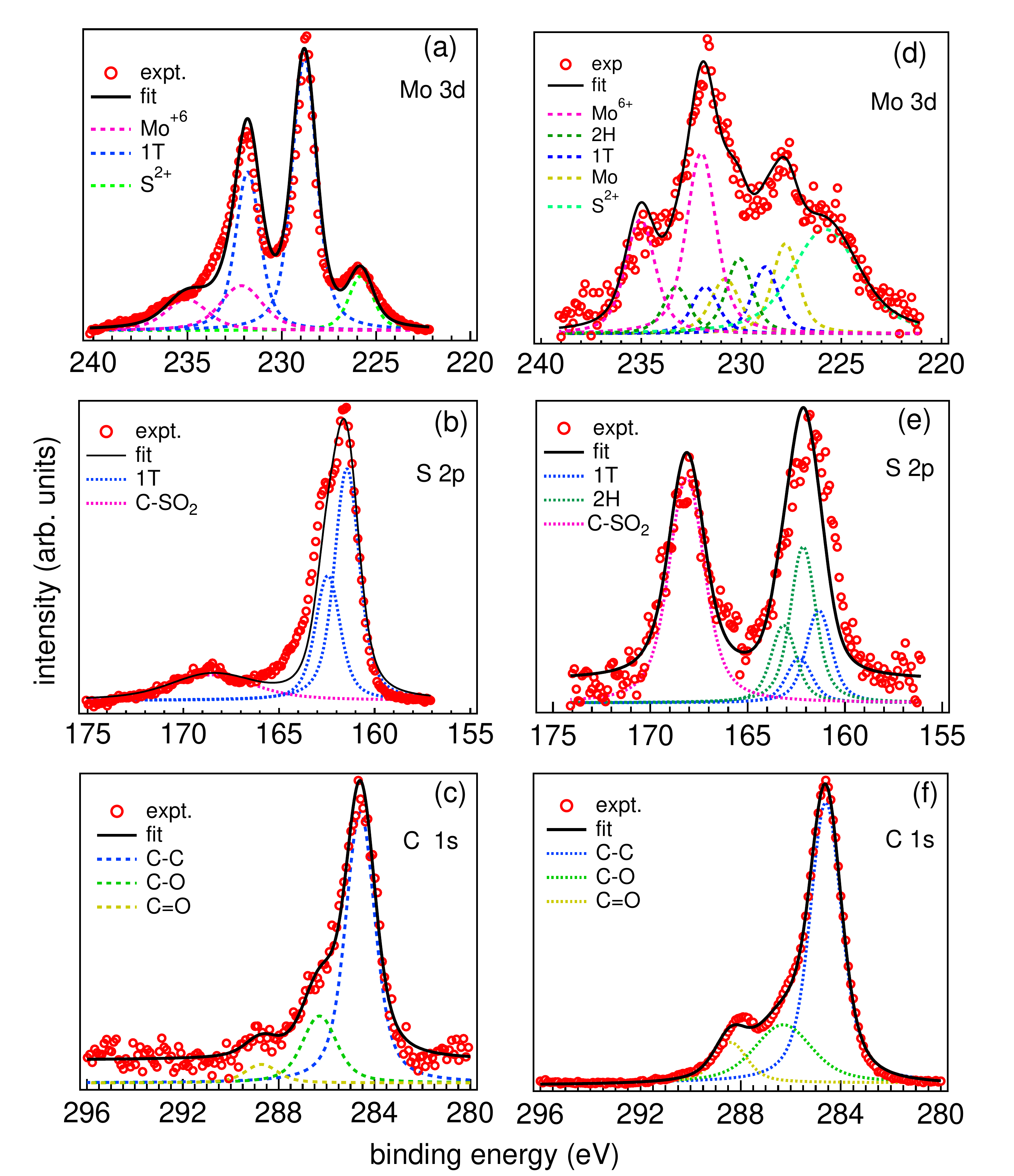}
\caption {The Mo 3$d$, S 2$p$ and C 1$s$ core-level XPS spectra of Al--MoS$_2$@rGO pristine sample (a--c) as well as powder extracted from electrode after long cycling (d--f), respectively.}
 \label{XPS}
\end{figure}

We use x-ray photoelectron spectroscopy (XPS) to investigate the electronic structure of Al--MoS$_2$@rGO pristine sample as well as after long cycling. In Fig.~\ref{XPS}(a), the Mo 3$d$ core-level of pristine sample consists of two spin-orbit coupled peaks 3$d$$_{5/2}$ and 3$d$$_{3/2}$ at the binding energies around 228.8~eV and 231.8~eV, respectively, which are consistent and confirm the presence of Mo in 4+ oxidation state \cite{4, EIS2, Yu_NC}. The de-convoluted peaks that detected around 232.1~eV and 235~eV could be related to MoO$_3$ or MoO$_4$$^{2-}$, which are found during the decomposition of Mo precursor (NH$_4$)$_6$Mo$_7$O$_{24}$·4H$_2$O in the hydrothermal synthesis \cite{Zou_RP}. Also, we can clearly observe S 2$s$ core-level at binding energy (BE) of 225.8~eV, which originate from the sulfide species \cite{L_ASS}. In case of the S 2$p$ spectrum of pristine sample, as shown in Fig.~\ref{XPS}(b), there are two intense peaks at BEs around 161.4 and 162.4~eV denoting the S 2$p_{3/2}$ and S 2$p_{1/2}$ doublet, respectively, of the S$^{2-}$ oxidation state. The peak located at 168.5~eV can be assigned to the sulfate groups \cite{LutzANM22}. The C 1$s$ core-level of pristine sample is displayed in Fig.~\ref{XPS}(c) where the most intense peak at 284.6~eV indicates the C-C boning, and the peaks located at 286~eV and 288.7~eV can be ascribed to the oxygenated functional groups C-O and C=O, respectively, which show the intercalative mixing of rGO and MoS$_2$ \cite{Zou_RP}. 

More importantly, we perform the XPS measurements ({\it ex-situ}) of anode powder that was extracted from the electrode after long cycling, which reveal the presence of both 1T and 2H phases of MoS$_2$ due to the transition reaction mechanism in 1T--MoS$_2$ after long cycling. Note that the Mo 3$d$ orbitals of trigonal prismatically co-ordinated 2H--MoS$2$ splits in to three degenerate energy levels forming (i) $d_{z^{2}}$, (ii) $d_{xy}$ + $d_{x^{2}-y^{2}}$ and (iii) $d_{yz}$ + $d_{xz}$ due to crystal field splitting and there is a finite energy gap of $\mathtt{\sim}$1~eV between the first two orbital sets. Moreover, the two Mo $d$ electrons fill the lower $d_{z^{2}}$ orbital, indicating the semi-conducting behavior of 2H--MoS$_2$. However, in case of 1T--MoS$_2$, there are two degenerate energy levels namely, (i) $d_{xy}$ + $d_{yz}$ + $d_{xz}$ and $d_{z^{2}}$ + $d_{x^2-y^2}$ due to the octahedral co-ordination and having two un-paired electrons in the lower (d$_{xy}$ + d$_{yz}$ + d$_{xz}$) orbitals make the 1T--MoS$_2$ metallic \cite{LeiAEM18, ChhowallaMRS15}. The more conductive nature of 1T--MoS$_2$ helps to enhance the electro-chemical kinetics due to fast electron transfer process as compared to the 2H--MoS$_2$. The intercalation of sodium-ions during cycling form the stable 2H phase and metastable 1T phase due to the charge transfer from alkali metal atoms to the MoS$_2$ nano-sheets. Fig.~\ref{XPS}(d) depicts the Mo 3$d$ spectrum of cycled material, where the co-existence of both 1T (octahedral) and 2H (trigonal) phases in the nano-sheets forming a 2D in-plane 1T/2H hetero-structure is clearly visible. The evolution of a peak at 227.8~eV gives a clear evidence of the presence of Mo 3$d_{5/2}$, which is formed during the conversion reaction of Na$_x$MoS$_2$  \cite{ZhuRSC21}. According to the composition, the cycled material consists of Na$_{2}$MoO$_{4}$, 1T--MoS$_2$ and 2H--MoS$_2$. Moreover, the Mo$^{6+}$ always exist in both pristine and cycled material indicating the formation of MoO$_{4}$$^{2-}$ where the Na$_2$MoO$_4$ acts as a protective sheath and provides stability during long cycling \cite{ZhuRSC21}. Further, the S 2$p$ core-level of cycled anode material is shown in Fig.~\ref{XPS}(e), which indicates the evolution of a pair of new peaks (at 162.2~eV and 163.2~eV for the doublet 2$p_{3/2}$ and S 2$p_{1/2}$, respectively). These components are consistent with the presence of 2H phase after long cycling and observed a shift of 0.8~eV towards higher BE as compared to the peaks of 1T--MoS$_2$ (at 161.4  and 162.4~eV) phase \cite{LiNat18}. Moreover, the Al 2$p$ core-level of cycled material is found to be at 74.5~eV BE, see Fig.~4 of \cite{SI}, which indicates the bonding between Al and S in the anode material. Fig.~\ref{XPS}(f) shows the C 1$s$ core-level spectrum of the cycled anode material where there is no shift in the peak positions as compared to the pristine material, indicating no change in oxidation state and stable nature of rGO. The evolution of mixed 1T/2H phase after long cycling affects the electrochemical performance owing to the formation of less conductive 2H--MoS$_2$ phase \cite{phase}, which is compatible with the long cycling results in Fig.~\ref{fig3}. 

\textbf{3.8 ~Theoretical results and discussion:}

\begin{figure}
\includegraphics[width=3.2in]{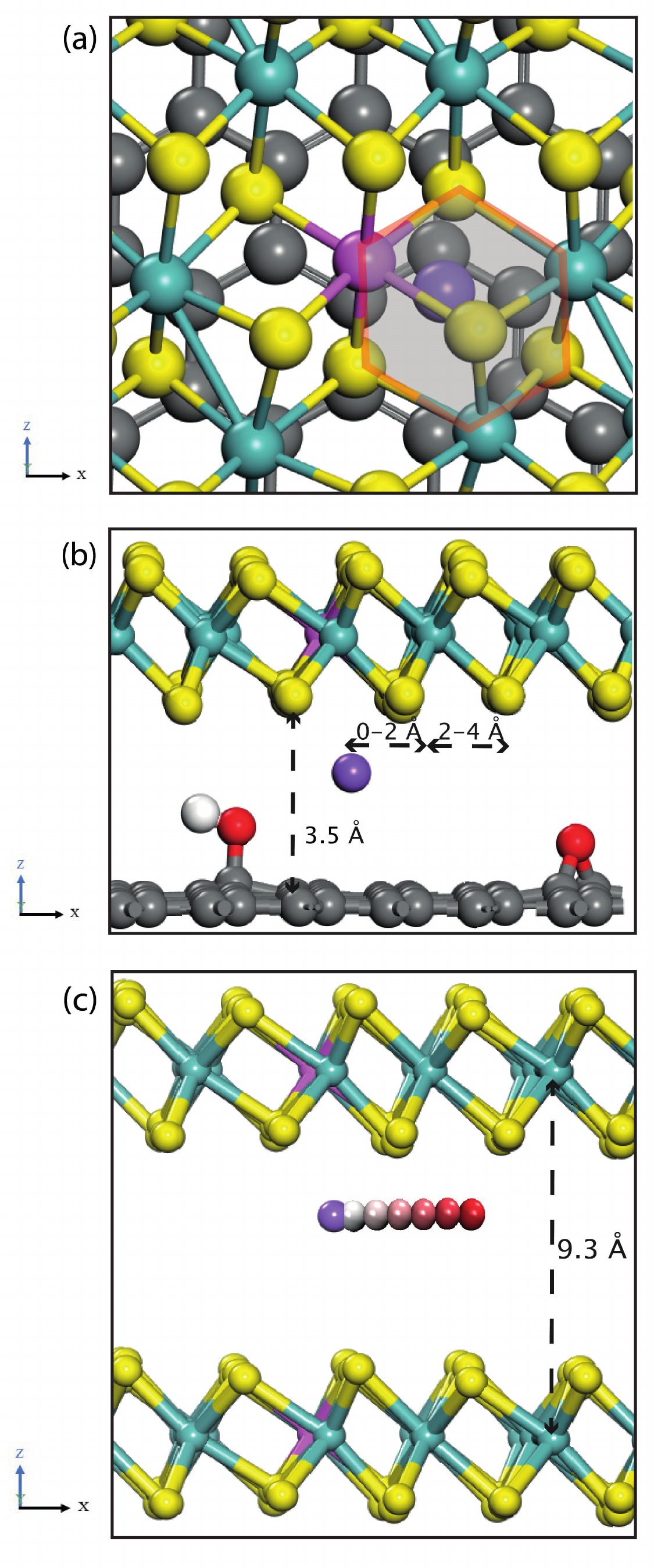}
\caption {(a) The top view of sodium atom (blue) at 2H site in the host Al (purple) doped 1T--MoS$_2$@rGO structure where grey hexagonal block shows the coordination of sodium with MoS$_2$ having Mo (cyan), S (yellow), C (black), O (red) and H (white). (b) The front view of optimized structure of sodium-ion intercalation in layered 1T phase of Al--MoS$_2$@rGO interface, where the sodium-ion activation barrier is visible. (c) The sodium-ion migration (blue-white-red) pathways in 1T Al--MoS$_2$@rGO interlayer and the activation barrier of sodium-ion along the migration path.}
\label{Fig9}
\end{figure}

Finally, using density functional theory (DFT) we investigate the Na-ion migration in 1T phase of Al--MoS$_2$@rGO interface and Al--MoS$_2$--MoS$_2$ interlayer host structures. We consider a monolayer MoS$_2$ supercell by cleaving a layer from the primitive unit cell of MoS$_2$. The unit cell of 1T-MoS$_2$ has one formula unit with hexa-coordinate geometry corresponding to the $\mathrm{P}\overline{3}\mathrm{~m}1$ space group \cite{S10}. Here the boing of Al and S is considered based on the XPS measurements, which are consistent with the observation in ref.~\cite{XPS_Al2S3} that the Al bonded with S atoms when doped in MoS$_2$. Also, the DFT calculations have further confirmed a greater stability of Al doping at the Mo site, which is bonded with the S atoms \cite{ZhangCPL19, LuoCPL16}, as shown in Fig.~\ref{Fig9}(a). The rGO is modelled in a supercell with approximately similar lattice parameters to that of MoS$_2$ supercell. The optimised structures are considered for the formation of layered 1T-MoS$_2$ at rGO layer with the initial interspace distance (distance between a carbon atom of rGO and its nearest S atom of MoS$_2$ layer) of 3.5~\AA. Here, the layering of Al--MoS$_2$ bulk and rGO is formed as an extended layer, one above the other with slight difference in the lattice parameters of individual structures but, the binding energy of this system is comparable with that of MoS$_2$/graphene interface, after the geometry optimisation of the structure \cite{S4}. This initial interspace distance is used for the optimization of 1T--MoS$_2$@rGO as the stable configuration, which is similar to the one obtained in MoS$_2$/graphene structure \cite{S11}. In order to introduce the sodium-atom in the structure, four different sites are available as explored by Massaro and co-workers on sodium-ion intercalation in the host MoS$_2$/graphene interface, which were explained as hollow-edge, hollow-hollow, top-edge and top-hollow where the sodium could be introduced between MoS$_2$-graphene layer. Interestingly, the calculated sodium-ion intercalation energy ($\Delta$$E$$_{\rm int}$) is to be the same and all these sites are considered stable sites for sodium atoms \cite{S4}. In the present work, the sodium-ion is introduced on structural optimization of Al--MoS$_2$/rGO composite at the hollow-hollow site, which comprised of one of the four sites mentioned above, see Fig.~\ref{Fig9}(a). The intercalation energy of sodium-ion at these sites is calculated by using the equation [$\Delta$$E$$_{\rm int}$ = $E$$_{\rm Na@site}$ -- $E$$_{\rm site}$ -- $E$$_{\rm Na}$]. The sodium sites have intercalation energy of -87.61~kJ/mol (= --0.908~eV), which is comparable to that of MoS$_2$/graphene and considered a stable site for sodium atom in the host Al-MoS$_2$@rGO structure. The optimized structure of 1T Al--MoS$_2$@rGO interface is shown in Fig.~\ref{Fig9}(b). 

Moreover, the sodium-ion migration is performed along the $x-$axis for migration distances corresponding to 0--2~\AA~and 2--4~\AA~where the reference is considered as the initial ($x=$ 0) position of sodium-ions at the hollow-hollow site, as represented in Fig.~\ref{Fig9}(a). An estimate for sodium-ion diffusivity is obtained using an Arrhenius type fit \cite{S12} $\mathrm{D}_{\mathrm{Na^+}}=\mathrm{D}_{0} \exp \left(-E_{a c t}/k_{B} T\right)$, where D$_0$ = $a^2$$v$, $a=$ sodium-ion migration length, $v=$ phonon frequency ($\sim$10$^{13}$ Hz) \cite{S9}, E$_{act}$ = activation energy, k$_B$ is the Boltzmann constant and $T$ is the temperature. At 300~K (k$_B$T = 0.0265 eV), the activation energies of sodium-ion are calculated as $\sim$57.9~kJ/mol and $\sim$40.3~kJ/mol in 1T Al--MoS$_2$@rGO interface for $a=$ 0--2~\AA~ and $a=$ 2--4~\AA, respectively. For the comparison, we have shown the activation energy values of un-doped 1T phase of MoS$_2$/rGO composite, which are found to be $\sim$68.3~kJ/mol and $\sim$50.2~kJ/mol for $a=$ 0--2~\AA~ and $a=$ 2--4~\AA, respectively. This suggests that the Al doped 1T phase have lower energy barrier as compared to the un-doped system. From the calculated activation energies using DFT, the corresponding diffusion coefficient of sodium-ions (D$_{Na^+}$) are estimated as $\sim$10$^{-13}$ and $\sim$10$^{-10}$ cm$^2$/s, respectively, which are comparable with the experimental values and indicate the faster diffusion of sodium-ion in the doped system. Note that the DFT simulations suggest that the sodium-ion activation energies vary from $\sim$40.3~kJ/mol to $\sim$57.9~kJ/mol with respect to the sodium-ion migration in 1T phase Al--MoS$_2$@rGO interface system and corresponding fits for D$_{Na^+}$ may vary from $\sim$10$^{-10}$ cm$^2$/s to $\sim$10$^{-13}$ cm$^2$/s. For the second model, the 1T phase Al--MoS$_2$--MoS$_2$ interlayer, a similar procedure is followed to study the sodium-ion migration. The interlayer distance (between two Mo atoms along $c-$axis) of 9.3~\AA~ is considered for DFT calculations, similar to the distances indicated in the experiment. The activation energies for sodium-ion migration are estimated as $\sim$49.8~kJ/mol to $\sim$59.6 kJ/mol, respectively for the aforementioned initial migration distances, i.e., from $a=$ 0--2~\AA~ to $a=$ 2--4~\AA. Fig.~\ref{Fig9}(c) shows the pathways of sodium-ion along the $x-$direction in the structure. The corresponding D$_{Na^+}$ values are estimated in the range of $\sim$10$^{-11}$--10$^{-13}$ cm$^2$/s, which suggests that sodium-ion transport is enhanced when moving from bulk to 2D layered structure \cite{S15}.  \\

\section{\noindent ~Conclusion}

The 1T phase of 5\% Al doped MoS$_2$@rGO was synthesized successfully through the hydrothermal route, which resulted in the larger interlayer distance of around 0.93~nm and the morphological studies confer the 3-D nano-flower-like architecture of the composite. This composite material as an anode for SIBs has displayed the first discharge specific capacity of about 764 mAhg$^{-1}$ with a reversible capacity retention of 383~mAhg$^{-1}$ over 10 cycles at a current density of 50 mAg$^{-1}$. We observed capacity retention of 86\%, 68\%, 58\%, and 45\% at current densities of 100, 300, 500, and 1000~mAg$^{-1}$ respectively, when the composite electrode was cycled over 200 cycles for stability test. The CV, EIS, and GITT analyses are performed to explore the sodium ion storage and diffusion mechanism in the composite host material. The calculated diffusion coefficient was found to be in the range of 10$^{−10}$--10$^{−12}$ cm$^2$s$^{−1}$ depending on the respective parameter of the experiment. The enhanced electrochemical performance of 1T phase Al--MoS$_2$@rGO composite is attributed to the collaborative effect of enlarged interlayer distance and larger electronic conductivity, which facilitate the storage of sodium ions and electron transportation. These results further motivate to explore 1T phase of 2D layered material as anode for sodium-ion batteries. The {\it ex-situ} XPS study confirm the presence of both 1T and 2H phases in the anode material, which reflects the small capacity fading after long cycling. The sodium-ion activation energies for migration in the layered 1T phase Al--MoS$_2$@rGO interface and Al--MoS$_2$--MoS$_2$ interlayer host structure are estimated using DFT. 

\section{\noindent ~Acknowledgments}

This work and setting-up the research facilities for sodium-ion battery project at IIT Delhi are financially supported from DST through ``DST--IIT Delhi Energy Storage Platform on Batteries" (project no. DST/TMD/MECSP/2K17/07) and from SERB-DST through core research grant (file no.: CRG/2020/003436). Manish, Jayashree and Jagdees thank MHRD, UGC, and DST (DST/TMD/MECSP/2K17/07), respectively for the fellowship. RSD and JKC thank IIT Delhi (IRD no. MI02186G) and NCTU, Taiwan, for the financial support through M-FIRP project. We thank IIT Delhi for providing research facilities for sample characterization (XRD and Raman scattering at the physics department, and FE-SEM and HR-TEM at CRF). The authors would like to acknowledge high performance computing (HPC) facility for all the computational work at IIT Delhi.

\end{document}